# Formation of highly stable interfacial nitrogen gas hydrate overlayers under ambient conditions

Chung-Kai Fang, Cheng-Hao Chuang, Chih-Wen Yang, Zheng-Rong Guo, Wei-Hao Hsu, Chia-Hsin Wang, and Ing-Shouh Hwang*

Institute of Physics, Academia Sinica, Nankang, Taipei 11529, Taiwan

Department of Physics, Tamkang University, Tamsui, New Taipei City 251301, Taiwan

National Synchrotron Radiation Research Center, 101 Hsin-Ann Road, Hsinchu 300092, Taiwan

Correspondence and requests for materials should be addressed to I.-S.H. (email: ishwang@phys.sinica.edu.tw).

**Abstract**

Surfaces (interfaces) dictate many physical and chemical properties of solid materials and adsorbates considerably affect these properties. Nitrogen molecules, which are the most abundant constituent in ambient air, are considered to be inert. Our study combining atomic force microscopy (AFM), X-ray photoemission spectroscopy (XPS), and thermal desorption spectroscopy (TDS) revealed that nitrogen and water molecules can self-assemble into two-dimensional domains, forming ordered stripe structures on graphitic surfaces in both water and ambient air. The stripe structures of this study were composed of approximately 90% and 10% water and nitrogen molecules, respectively, and survived in ultra-high vacuum (UHV) conditions at temperatures up to approximately 350 K. Because pure water molecules completely desorb from graphitic surfaces in a UHV at temperatures lower than 200 K, our results indicate that the incorporation of nitrogen molecules substantially enhanced the stability of the crystalline water hydrogen bonding network. Additional studies on interfacial gas hydrates can provide deeper insight into the mechanisms underlying formation of gas hydrates.

Clathrate hydrates, or gas hydrates, are crystalline solids in which water molecules form cages containing small non-polar gas molecules. Typically, gas hydrates are unstable under ambient conditions; they form under pressures substantially higher than ambient pressure and at temperatures considerably below room temperature (RT). Many experimental studies have demonstrated that hydrophobic solid particles play a role in promoting gas hydrate formation[1-5]. Some studies have indicated that hydrophobic solid surfaces shift the equilibrium condition for gas hydrate formation to lower pressures and higher temperatures[4]. In addition, nucleation and growth of methane hydrates in the confined nanospace of activated carbons occur under milder conditions and with faster kinetics than those observed in nature[6]. To date, the understanding of the thermodynamics and kinetics involved in the formation of gas hydrates, particularly at the interfaces between water and hydrophobic solids, remains inadequate. Enhanced understanding of these processes can considerably improve the ability to control gas hydrate formation. In the current study, we identified a specific type of nitrogen gas hydrate layer that forms on graphitic surfaces, which are mildly hydrophobic substrates, in water under ambient conditions and in ambient air with a certain level of humidity.

This nitrogen gas hydrate layer was initially observed through atomic force microscopy (AFM) and appeared as two-dimensional (2D) domains of ordered row-like (or stripe) structures with a height of approximately 0.5 nm and row separations of 4 to 6 nm on graphitic surfaces. In 2012, our group reported the formation of stripe structures on highly ordered pyrolytic graphite (HOPG) in deionized water[7-10]; these stripe structures, aligning along the zig-zag direction (Fig. S1), gradually nucleated and grew in lateral size over several hours. These structures were initially proposed to form though the adsorption of dissolved nitrogen molecules at the HOPG-water interface, because a nitrogen gas environment possibly promoted nucleation and growth of the stripe structures[7,8,10]. Subsequently, these stripe domains are determined to play a crucial role in the formation of surface nanobubbles; they have often been seen at the nanobubble-water-graphite contact line and pin the lateral movement of surface nanobubbles[9,11]. On the basis of AFM observations of the nucleation process of surface nanobubbles, our group proposed that the stripe structures represent an interfacial gas hydrate layer[11]. Similar stripe domains were later observed on graphene (covered on mica substrates) in water saturated with air gas[12,13]. In 2018, Foster et al. noted similar stripe domains on HOPG in ultrapure water, with these domains identified using AFM, and indicated that the structures were formed because of the self-assembly of water and methanol molecules at the HOPG–water interface[14]; they proposed that a catalytic conversion of dissolved carbon dioxide and water into methanol had occurred at HOPG step edges. Several other research groups from various countries have reported similar findings of stripe structures on HOPG in water[15,16]. Seibert et al. reported that the stripe domains readily form on HOPG when standard plastic syringes are used to insert water into the AFM instrument. However, they did not observe such stripes when using clean glass syringes[17] These findings suggest that the stripe domains result from the presence of chemical species in plastic syringes. In addition to stripe structures similar to those reported on HOPG in water, Seibert et al. observed domains with stripes that did not align along the zig-zag or arm-chair direction of HOPG substrates and domains with small stripe spacing (~2 nm)[17].

Domains of similar stripe structures have been reported for aged graphitic samples, including HOPG and graphene samples exposed to ambient air for several days[18-22]. Through AFM, these self-assembled stripe structures were identified as responsible for the anisotropic friction domains present on graphene[20,22]. Similar anisotropic friction domains or stripe structures have been noted on several van der Waals (vdW) materials, including hexagonal boron nitride[20], molybdenum disulfide[23–25] and tungsten disulfide[26]. Airborne hydrocarbons, which are common air pollutants in laboratories, were proposed to be the source of these self-assembled stripe structures[20–22]. In 2022, using low-temperature scanning tunneling microscopy (STM) and infrared spectroscopy, Pálinkás et al. determined that these stripe structures result from the adsorption and self-assembly of mid-length normal alkanes of 20–26 carbon atoms present in the environment[27]. Moreover, they observed that molecules lie parallel to the HOPG zig-zag axis and that the stripes are parallel to the arm-chair direction. Because of the low concentration of mid-length normal alkanes in ambient air, these stripe structures typically form after a few days of ambient exposure, with no formation being observed within 24 hr of measurement[27]. For many years, it was generally believed that the stripe structures that form on HOPG in water had the same origin as those formed on vdW materials after exposure to ambient air for days because of the similarity of AFM images of these structures. However, these structures have major differences. The stripe structures that form on HOPG in water align parallel to the zig-zag direction[7–10], which differs from the arm-chair direction observed for stripe structures that form on aged graphitic samples[20,21,27]. In addition, the stripe structures that form on HOPG in water are fragile and can be easily destroyed by the AFM tip if the imaging force is not sufficiently small[7–9]. The stripe structures on aged graphene or vdW materials are strong enough to withstand AFM imaging forces, although the stripe direction may be reoriented under strong force[20,22,27]. Because the concentration of mid-length normal alkanes in pure water is extremely low (well below 1 ppb), the formation of stripe structures on freshly cleaved HOPG or freshly prepared graphene within a few hours of water deposition likely have different causes. In the current study, we present evidence demonstrating that the stripe structures that form on HOPG in water are nitrogen gas hydrate layers.

Although stripe structures may form on aged HOPG[22,27], they are much less frequently observed on aged HOPG than on aged graphene[22,28]. A sprinkle of graphene oxide nanoflakes (nanoGOs) can result in the formation of stripe structures on HOPG under ambient air conditions with humidity levels above 15%[28]. Because superhydrophilic nanoGOs can condense water from ambient air and seed the formation of the stripe structure on HOPG, these stripe structures are considered as RT ice overlayers[28]. In the current study, we determined that the stripe structures that form on HOPG after the deposition of nanoGOs are also nitrogen gas hydrate layer. We mainly studied two types of samples: $HOPG_{water}$, which are HOPG samples with the stripe structures that form in water, and $HOPG_{nanoGOs}$, which are HOPG samples with the stripe structures that form after the deposition of nanoGOs under ambient air conditions. We also performed measurements on freshly cleaved HOPG ($HOPG_{fresh}$), with the findings for the $HOPG_{fresh}$ compared with those of the aforementioned types of samples. We employed AFM, X-ray photoemission spectroscopy (XPS), and thermal desorption spectroscopy (TDS). Unless otherwise specified, all experiments were conducted at RT (22-24 °C). We used AFM to obtain surface topography of the stripe structures at nanometer resolution. XPS and TDS have excellent chemical sensitivity of solid surfaces; however, XPS and TDS measurements should be conducted under ultra-high vacuum (UHV) conditions. Using AFM, we determined that the stripe structures that formed on $HOPG_{water}$ and those that formed on $HOPG_{nanoGOs}$ survived in vacuum conditions. XPS revealed a strong signal for oxygen K-edge and a smaller signal for nitrogen K-edge on the $HOPG_{water}$ and the $HOPG_{nanoGOs}$ samples. TDS, which is based on mass spectroscopy, detected a significant peak in desorbed water (mass 18) and a smaller peak in $N_2$ (mass 28) at temperatures within the range of 70 °C and 100 °C for both types of samples. Neither methanol nor $O_2$ (mass 32) was detected. These findings indicate that the stripe structures that formed on the $HOPG_{water}$ and $HOPG_{nanoGOs}$ samples are nitrogen gas hydrate overlayer.

## Results

**AFM: stripe structures survive in vacuum conditions**

Fig. 1a and b presents the formation of a domain of stripe structure on $HOPG_{water}$ and the survival of the stripe domain after water removal, respectively. Fig. 1c presents another $HOPG_{water}$ sample exhibiting the presence of several domains of stripe structures in water. After we removed the water, we immediately transferred the sample into a vacuum chamber and maintain the pressure at approximately $1\times10^{-9}$ torr overnight. Subsequently, we moved the HOPG sample to ambient air to complete AFM imaging of the regions that we had previously observed in water. Most of the stripe domains remained intact (Fig. 1d, e), indicating that the stripe structures were stable in vacuum. Thus XPS and TDS can be applied for chemical analysis of the stripe structures.

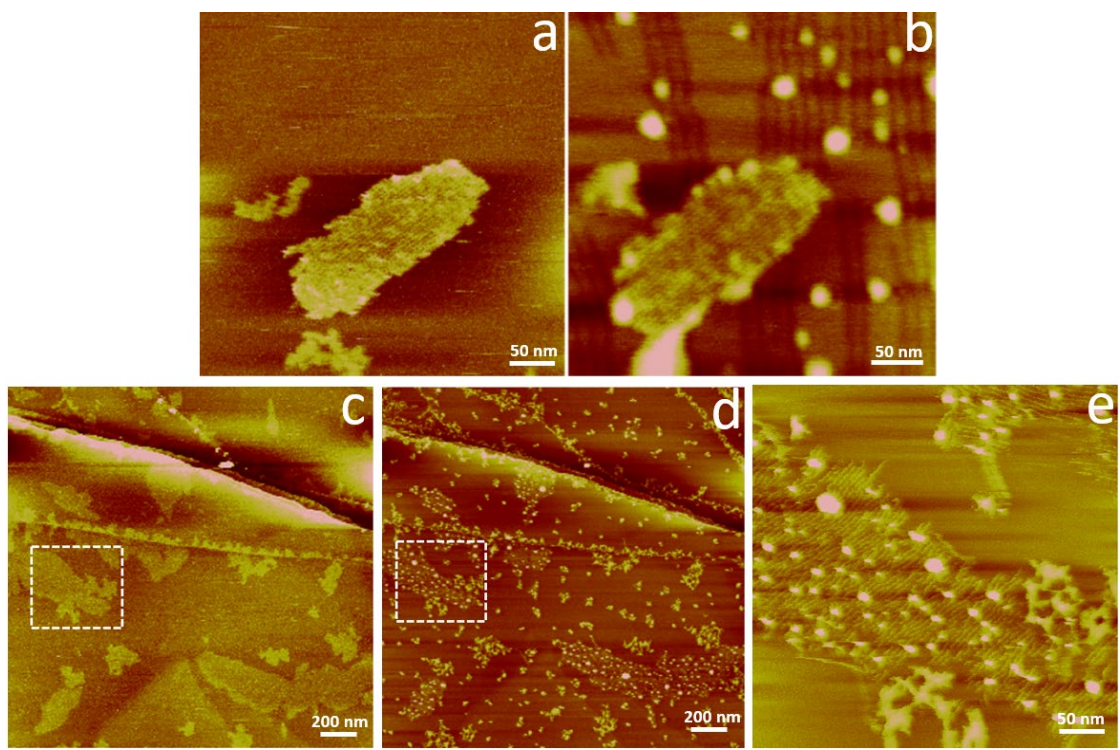

**Fig. 1 Stripe domains formed on $HOPG_{water}$ survive after water removal and storage in**

**vacuum conditions. a** AFM height image illustrating a domain of stripe structures on HOPG in deionized water. **b** Height image showing the same region as that in (**a**) in ambient air after water removal. Additional particles and structures of unknown origin were observed after the water removal. **c** Height image of another sample of HOPG$_{water}$ in deionized water. **d** Height image presenting the same region same as in (**c**) after water removal and storage in vacuum conditions. The sample was imaged after it was moved to air. The white dashed box outlines a region for comparison between (**c**) and (**d**). **e** Stripe structures evident in a high-resolution image of the region outlined in the white dashed box in (**d**).

Fig. 2a presents a height image of an HOPG$_{nanoGOs}$ sample acquired in ambient air. The high relative humidity (RH, 60%–80 %) in our lab led to the formation of one layer of stripe structures across the entire surface and the formation of the second layer on some regions. Although the stripe structures of the second layer are evident in the height image (Fig. 2a), those of the first layer are more easily discernible in the stiffness map (Fig. 2b). The stripe structures tended to form elongated domains, with their long axis parallel to the row orientation. Bright particles are super-hydrophilic nanoGOs[28], which condense water from ambient air. These particles tended to appear at the boundaries of domains with different row orientations. Point defects tend to segregate at domain boundaries and can pin domain walls[29]. The HOPG$_{nanoGOs}$ sample was stored in a vacuum chamber at approximately $1\times10^{-9}$ torr for 20 hr before it was moved to ambient air for AFM imaging. Although some second-layer domains disappeared, nearly every first-layer domains remained intact (Fig. 2c,d). In addition, most nanoGO nanoparticles remained in their original positions. These observations indicated that most of the stripe structures that formed on HOPG$_{nanoGOs}$ survived in vacuum conditions.

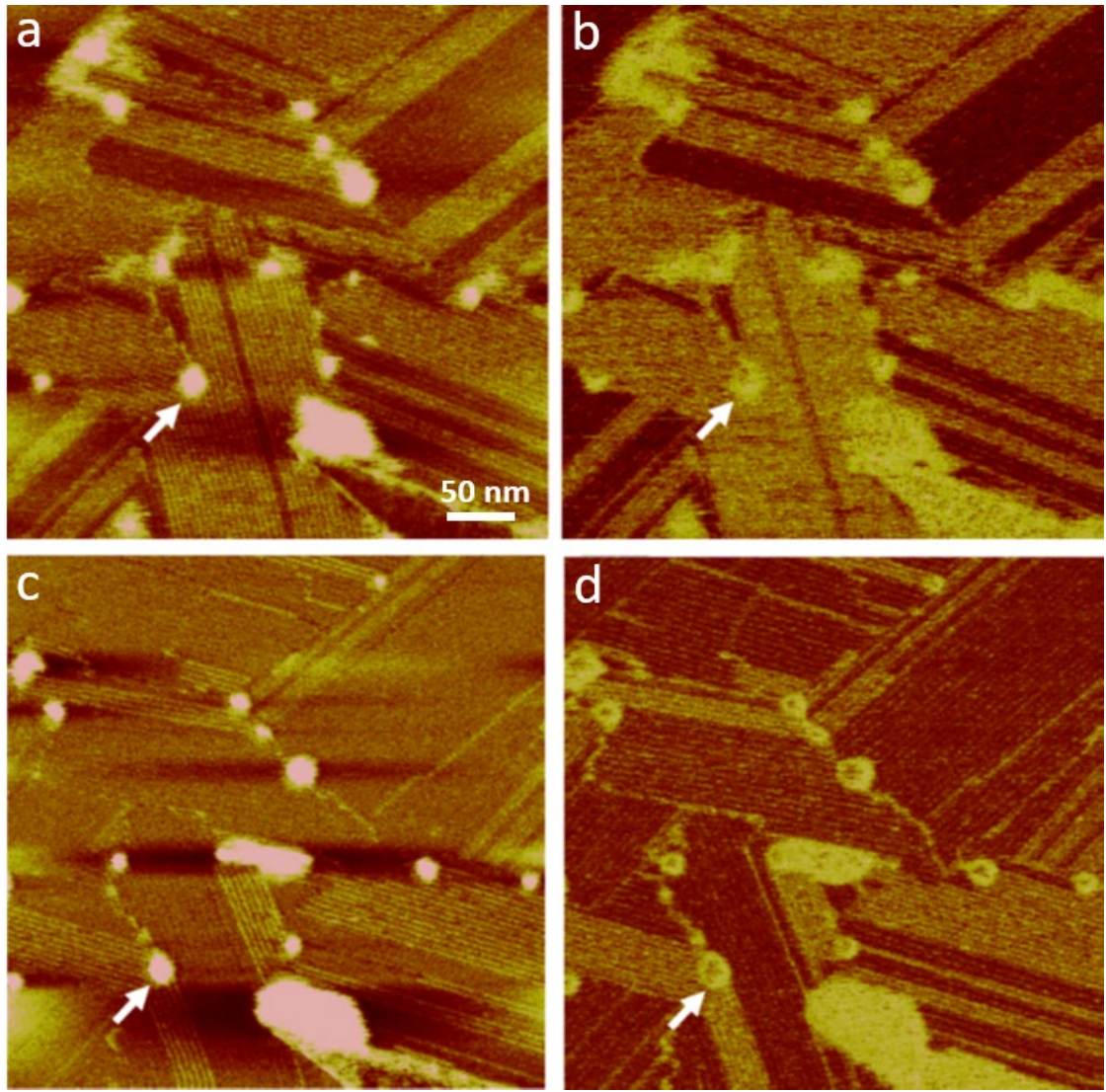

**Fig. 2 Most stripe structures formed on HOPG$_{nanoGOs}$ survive in vacuum conditions. a** and **b** respectively present the height and stiffness images of stripe structures that formed under ambient air (RH= 60-80%). The nanoGOs appear as nanoparticles with bright protrusions in the height images. **c** and **d** present the height and stiffness images of approximately the same region as that presented in **a** and **b** after the sample was placed in vacuum and moved to ambient air. The nanoGO particles indicated with white arrows serve as reference points for comparing these two sets of data. Note that the stiffness maps may not correctly represent the stiffness values of surfaces because sample structures are extremely thin.

The stripe structures that formed on the surface of HOPG$_{nanoGOs}$ gradually disappeared after undergoing annealing in ambient air (Fig. 3). Fig. 3a presents a height image of the stripe structures on an HOPG$_{nanoGOs}$ at RT. Although the stripe structures of the second layer are evident in this image (Fig. 3a), those of the first layer can be more easily discerned in the stiffness map (Fig. 3b). After the sample was annealed at approximately 70°C on a hot plate for approximately 15 min, most of the stripe structures remained (Fig. 3c, d). However, some second-layer stripe domains disappeared, and for some, their stripe orientation changed. In addition, some nanoGO particles were displaced. The second-layer stripe structures nearly disappeared after annealing at 90°C for 15 min. (Fig. 3e, f), and most of the nanoGO particles disappeared from the imaged area. Although the stripe domains on the first layer remained, the domain orientation changed, and the domain size increased. The stripe structures and nanoGOs in this region completely disappeared after undergoing annealing at 135 °C for 15 min (Fig. 3g, h).

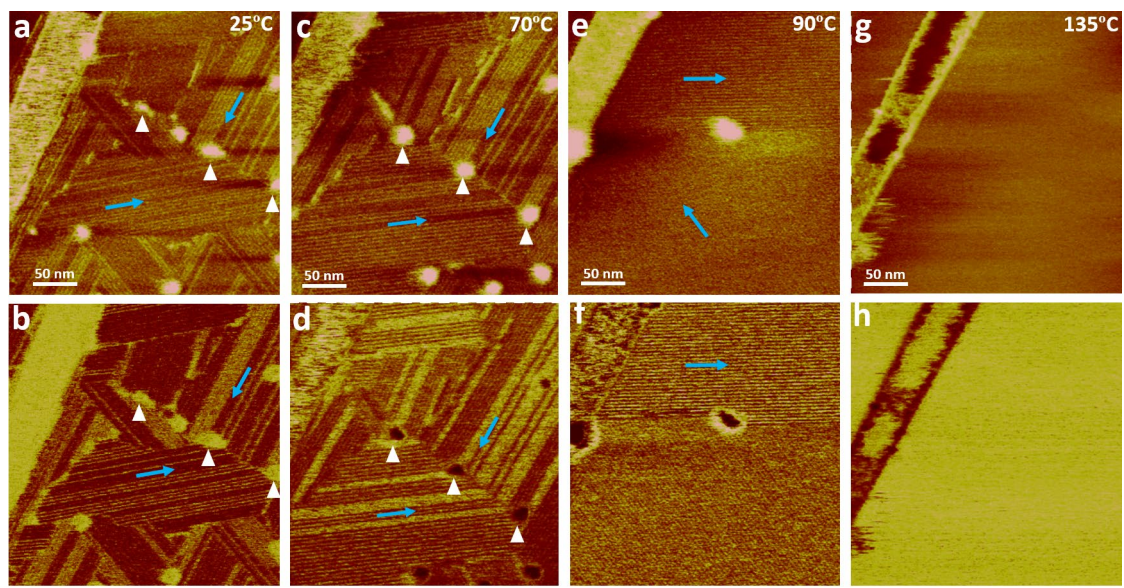

**Fig. 3 Thermal desorption of stripe structures formed on HOPG$_{nanoGOs}$ after annealing in air. a**, **c**, **e**, and **g** present height images. The temperature of the sample in each image is indicated in the upper-right corner. The images were acquired after the sample was cooled down to RT. **b** Corresponding stiffness map acquired along with **a**. **d, f,** and **h** present corresponding adhesion maps acquired along with **c**, **e**, and **g**, respectively. A few nanoGO nanoparticles are marked with white arrowheads; blue arrows indicate the row orientation of the stripe domains.

**XPS results**

We performed XPS measurements on HOPG$_{water}$, HOPG$_{nanoGOs}$, and HOPG$_{fresh}$ samples over 4 years, with the experiments being repeated more than 10 times. The results remained consistent across all measurements. Fig. 4 presents the results of the wide-range scans covering C, O, and N 1s core levels collected for the HOPG$_{water}$ (Fig. 4a), HOPG$_{nanoGOs}$ (Fig. 4b), and HOPG$_{fresh}$ (control; Fig. 4c) samples. In addition to the strong C$_{1s}$ signal at the binding energy (BE) of approximately 285 eV, which corresponds to that of the HOPG substrate, a strong O$_{1s}$ signal at the BE of approximately 533 eV and a weak N$_{1s}$ signal at the BE of approximately 400 eV were detected for the samples exhibiting stripe structures on their surfaces. For the HOPG$_{fresh}$ samples, the N$_{1s}$ signal was absent, and the O$_{1s}$ signal was weaker than those for the HOPG$_{nanoGOs}$ and HOPG$_{water}$ samples (Fig. 4c). The O$_{1s}$ and N$_{1s}$ signals were stronger for HOPG$_{nanoGOs}$ than for HOPG$_{water}$ likely because of the higher coverage of stripe structures on HOPG$_{nanoGOs}$.

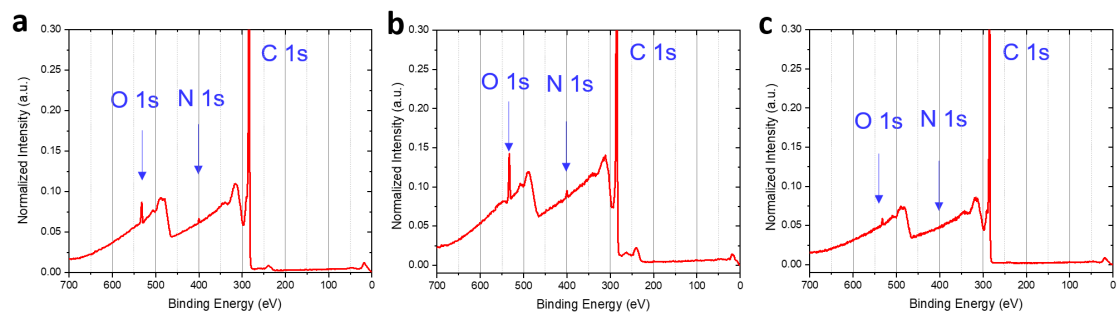

**Fig. 4 Typical XPS survey scans**. **a** HOPG$_{water}$. **b** HOPG$_{nanoGOs}$. **c** HOPG$_{fresh}$. The intensity of each scan was normalized by setting the peak of C 1s as 1.

Fig. 5 illustrates the high-resolution O$_{1s}$ and N$_{1s}$ spectra of the three types of samples. The O$_{1s}$ spectrum of HOPG$_{fresh}$ exhibited a single peak at the BE of 532.3 ± 0.2 eV (red dotted lines in Fig. 5a, b); the spectra of HOPG$_{water}$ and HOPG$_{nanoGOs}$ revealed a broader peak centered at a higher BE of approximately 533 eV (black line in Fig. 5 a, b). A study reported that the physical wetting of $H_2O$ molecules on HOPG and $O_2$ chemisorption on HOPG yielded an $O_{1s}$ peak at a BE of approximately 533 and 532 eV, respectively[30]. Additionally, the $O_{1s}$ peak of water ice on solid substrates was reported to have a BE between 532.8 and 533.4 eV[31-34]. On the basis of these spectral results, we attributed the small $O_{1s}$ signal on the $HOPG_{fresh}$ surface to the chemisorption of $H_2O$ or $O_2$ molecules on the step edges or other defective sites of the surfaces (red dotted lines in Fig. 5a, b). The presumed stripe structures on the $HOPG_{water}$ and $HOPG_{nanoGOs}$ surfaces were associated with $H_2O$ molecules. The $O_{1s}$ spectra for $HOPG_{water}$ and $HOPG_{nanoGOs}$ were similar, with the exception that the spectrum of $HOPG_{nanoGOs}$ was slightly wider than that of $HOPG_{water}$ due to the presence of nanoGOs.

The $N_{1s}$ spectra of $HOPG_{water}$ and $HOPG_{nanoGOs}$ were similar and exhibited peaks at the BE of approximately 400 eV (Fig. 5 c, d). However, the $N_{1s}$ spectra of $HOPG_{fresh}$ revealed a flat background (red dotted lines in Fig. 5 c, d), indicating the absence of nitrogen. The stripe structures on HOPG have never been studied using XPS. Previous studies using XPS to investigate the adsorption of $N_2$ molecules on carbon nanotubes[35], $TiO_2$ fiber[36], and Cr/W(110)[37] have revealed $N_{1s}$ spectra with BEs of 399.2, 400.1, and 400.0 eV, respectively. Thus, the finding of a peak of approximately 400.0 eV of the present study can be attributed to $N_2$ molecules in the stripe structures.

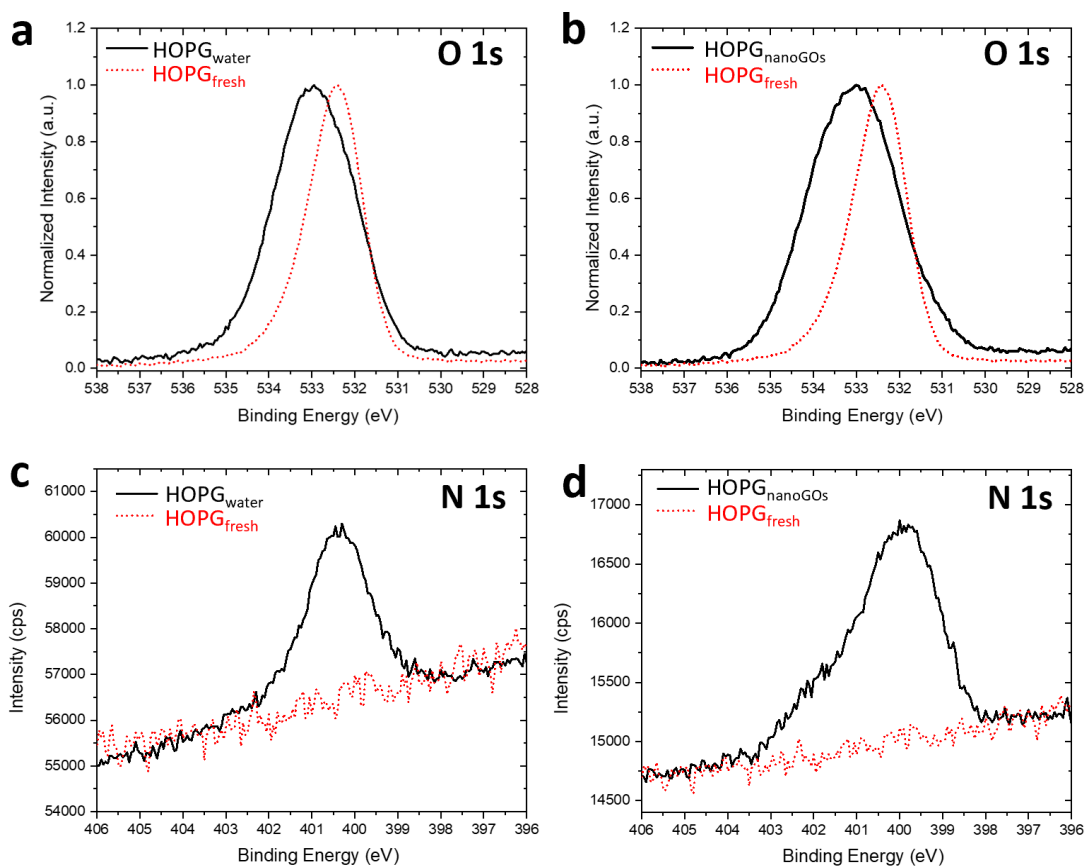

**Fig. 5 Typical XPS fine scans of $O_{1s}$ and $N_{1s}$ signals for the three types of samples. a** $O_{1s}$ spectra for $HOPG_{water}$ and $HOPG_{fresh}$. **b** $O_{1s}$ spectra for $HOPG_{nanoGOs}$ and $HOPG_{fresh}$. **c** $N_{1s}$ spectra for $HOPG_{water}$ and $HOPG_{fresh}$. **d** $N_{1s}$ spectra for $HOPG_{nanoGOs}$ and $HOPG_{fresh}$.

Pálinkás et al. performed XPS on aged graphite samples[27] and detected no $N_{1s}$ signal and a small $O_{1s}$ signal (<1%). The $O_{1s}$ and $N_{1s}$ spectra of our measurements indicate that the stripe structures on $HOPG_{water}$ and $HOPG_{nanoGOs}$ comprised water and $N_2$ molecules. To determine the molecular ratio in the stripe structures, we analyzed XPS fine scans of $O_{1s}$ and $N_{1s}$ signals. Our analysis revealed that the nitrogen gas hydrate was composed of 90% ± 4% $H_2O$ and 10% ± 4% $N_2$ (Supplementary Note 1).

### TDS results

We used three quadrupole mass spectrometers for our TDS measurements and obtained similar and consistent results on $HOPG_{fresh}$, $HOPG_{water}$, and $HOPG_{nanoGOs}$. Fig. 6 presents typical TDS of the three types of samples. For the $HOPG_{fresh}$ sample, we detected only a small peak of water (m/z = 18) at approximately 70 °C (Fig. 6a). This water might have originated from the sample holder because the holder was moved from a vacuum chamber to ambient air (humidity = 60%–80%) for approximately 5 min when the HOPG sample was being mounted. For the $HOPG_{water}$ sample, we detected strong desorption of water (m/z = 18) with a broad peak in the range of 60–100 °C (Fig. 6b). In addition, we detected smaller desorption peaks at m/z = 28 and 15; the peak at m/z = 28 might be due to the presence of $N_2$ and CO, and that at m/z = 15 might be due to the presence of the hydrocarbon fragment $CH_3$. The TDS spectra corresponding to the m/z of 28 and 15 were negligibly low for the $HOPG_{fresh}$ sample (Fig. 6a). To ensure that the detected water signals mainly originated from the HOPG surface rather than the sample holder, we prepared $HOPG_{water}$ in heavy water ($D_2O$) and performed TDS (Fig. 6c). Clear desorption signals of m/z of 18, 19, and 20 were detected (Supplementary Note 2). The signals of m/z of 19 and 20 were completely absent for the samples that were not prepared using heavy water, indicating that the stripe structures were mainly composed of water molecules.

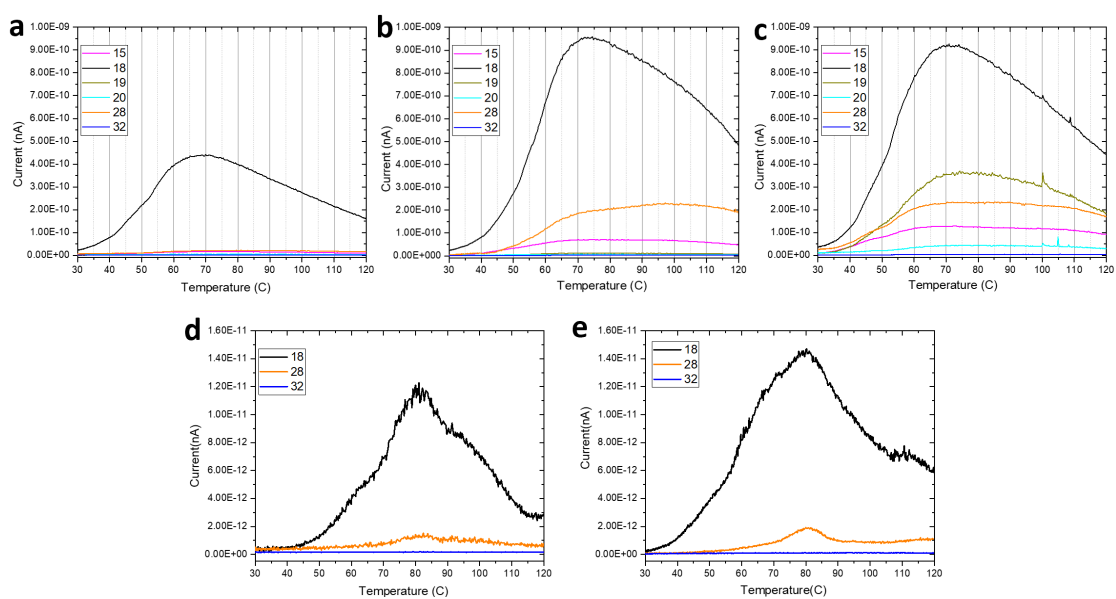

**Fig. 6 Thermal desorption spectra of HOPG surfaces. a** $HOPG_{fresh}$. **b** $HOPG_{water}$. **c** HOPG in $D_2O$. The spectra presented in **a**–**c** were measured using Pfeiffer PrismaPro® QMG 250. **d**

HOPG$_{water}$. **e** HOPG$_{nanoGOs}$. The spectra presented in **d** and **e** were measured using Pfeiffer Vacuum Prisma QMS 200. Different mass-to-charge ratio (m/z) values are indicated with different colors. The measured current is proportional to the desorption rate.

Fig. S2 presents the mass spectra measured on the HOPG$_{water}$ sample before heating and when it was heated at approximately 80 °C with m/z of 1 to 50. No desorption with m/z of 51 to 100 was detected (data not shown). The spectra suggest that the desorption peak at m/z of 28 can be attributed to N$_2$ molecules. We performed TDS on HOPG$_{water}$ more than 15 times and consistently obtained findings of spectra indicating strong desorption of water molecules and weaker desorption at m/z = 28. Fig. 6d, e presents the TDS spectra of HOPG$_{water}$ and HOPG$_{nanoGOs}$, respectively, which were determined using a mass spectrometer (Pfeiffer Vacuum Prisma QMS 200). We detected a strong desorption of m/z = 18 (water) and a weaker desorption of m/z = 28 at approximately 80 °C. The spectra were similar, indicating that the stripe structures of these two types of samples had similar chemical compositions. The desorption shapes presented in Fig. 6b and Fig. 6d differ slightly because of several factors, including the use of different spectrometers, variations in geometry and spacing between the head of the mass spectrometer and sample surface, and different heating rates for the samples. Overall, the general desorption behaviors of the same types of HOPG samples that were measured using different mass spectrometers were similar.

The signal at m/z = 32 remained considerably low during the heating process (Figs. 6 and S2), indicating that the stripe structures did not contain O$_2$ or CH$_3$OH. A weak hydrocarbon signal (m/z = 15) was detected. This signal was typically weaker than that at m/z = 28. This weak hydrocarbon signal might have been caused by the adsorption of hydrocarbon molecules when the HOPG sample and sample holder were exposed to ambient air for a few minutes.

The stripe structures resulting from the self-assembly of hydrocarbon contaminants on vdW materials disappeared after undergoing annealing at 200 °C for 1h[27]. This annealing temperature was considerably higher than the desorption temperature of the stripe structures on $HOPG_{water}$ and $HOPG_{nanoGOs}$, indicating that the stripe structures on $HOPG_{water}$ and $HOPG_{nanoGOs}$ did not result from hydrocarbon contaminants.

**Discussion**

Zheng et al. proposed that the stripe structures on $HOPG_{nanoGOs}$ are RT ice overlayers. Similarly, Zhao et al.[38] reported the formation of RT ice chains on suspended graphene and observed a unique electron diffraction pattern (in vacuum conditions) after the graphene sample was rinsed or sprayed with water under ambient conditions. These ice overlayers and ice chains remained stable in vacuum conditions at RT. This finding contradicts those of several experimental studies indicating that under UHV conditions, water molecules completely desorb from graphitic (HOPG or graphene) surfaces at temperatures lower than 200 K[39-42]. However, studies reporting RT ice overlayers and ice chains exposed their samples to water molecules under ambient conditions, whereas the latter studies have performed experiments in vacuum conditions. Our results reveal that the stripe structures on $HOPG_{nanoGOs}$ are nitrogen gas hydrate overlayers rather than RT ice overlayers. This finding indicates that nitrogen molecules, which have typically been considered to be inert, play a role in the formation of interfacial structures on solid surfaces under ambient conditions. Evidently, the incorporation of a small percentage of nitrogen molecules substantially enhanced the structural stability of the water hydrogen bonding network on the graphitic surfaces. The reported ice chains on graphene[38] might also be nitrogen gas hydrate overlayers because the direction of such chains is parallel to the crystal orientation of graphene (a zig–zag direction), which is identical to the stripe direction in $HOPG_{water}$.

Our observation that the stripe structures on HOPG$_{water}$ survive after water is removed (Fig. 1a, b) indicates that the nitrogen gas hydrate layer does not have a strong interaction with overlying liquid water. This finding aligns with that of a previous study demonstrating the formation of a hydrophobic water monolayer and nonwetting growth for subsequent crystalline water layers on Pt(111) at temperatures lower than 170 K[43]. The study proposed that molecules in the water monolayer form a fully coordinated surface with no dangling OH bonds or lone pair electrons. The water molecules in stripe structures on HOPG$_{water}$ might form hydrogen bonding networks with few or no dangling OH bonds or lone pair electrons on the surface, resulting in weak interactions of the structures with the overlying liquid water. Thus, studies should determine the details of the atomic structures and hydrogen bonding that occur in interfacial gas hydrate layers. Because current AFM imaging of stripe structures still do not have sufficient resolution to determine the atomic structures, researchers may be able to obtain more details of these stripe structures by using UHV-AFM at cryogenic temperatures; structural fluctuations[10] can be considerably minimized at low temperatures. Low-energy electron diffraction and other diffraction methods can be employed to examine stripe structures on HOPG. Additional experimental and theoretical investigations can be conducted to obtain details regarding atomic and molecular stripe structures and to identify reasons for the high stability of nitrogen gas hydrate layers on graphitic surfaces.

Whether the formation of nitrogen gas hydrate layers occurs on many other solid surfaces that come into contact with liquid water or are exposed to ambient air with humidity levels higher than a certain level remains unclear. Many studies have suggested such a possibility. Immobile surface nanobubbles have been observed on many different hydrophobic solid surfaces in water, and these surface nanobubbles might also be pinned by nitrogen gas hydrate layer formed at the three-phase nanobubble-water-surface contact line. A few studies reported the formation of a protruded circular rim at the perimeter of each surface nanobubble on polystyrene and a gradual increase in the rim height over several hours[44-46]; this phenomenon is similar to that of stripe structures being identified around surface nanobubbles on HOPG[11]. Further exploration of this topic by using a combination of AFM, XPS, and TDS on other systems can confirm this speculation. Gas hydrate overlayers forming on different solid surfaces would constitute an interfacial phenomenon that potentially affects the interfacial properties (e.g., wetting, adsorption, tribology, and chemical and electrochemical reactions) of solid surfaces under ambient conditions. Additionally, interfacial gas hydrate layers might serve as easily accessible systems that can be used to analyze the kinetics and formation mechanisms of gas hydrates.

## Methods

**Materials and Sample Preparation.** The HOPG samples (lateral sizes of 12 mm × 12 mm, Grade: ZYB) were provided by Momentive Technologies. The $HOPG_{fresh}$ sample was prepared by peeling off the top layer of a HOPG substrate with Scotch tape prior to each experiment. Water was purified using a Milli-Q system (Millipore, Boston) with a resistivity of 18.2 MΩ·cm. $D_2O$ (99.9 atom % D) was purchased from Sigma Aldrich. To prepare $HOPG_{water}$, an $HOPG_{fresh}$ substrate was placed in a liquid cell of AFM. Subsequently, deionized water was injected into the liquid cell. AFM imaging was conducted to monitor the growth of the stripe domains. For the XPS and TDS measurements, stripe structures were grown until they covered more than 50% of the surface. Subsequently, water was removed from the stripe structures, and they were placed in a desiccator pumped to approximately 0.1 atm. The desiccator was opened immediately before the measurements. The preparation method for nanoGOs has been detailed in a previous study[47]. $HOPG_{nanoGOs}$ were prepared by depositing a water solution of nanoGOs onto a HOPG$_{fresh}$ surface. Approximately 3 min later, all excess solution was blotted off the surface. The sample was dried on a hot plate at approximately 80 °C for 15 min.

**AFM.** Measurements were performed using a Bruker AXS Multimode NanoScope V equipped with a commercial liquid cell tip holder. We used the peak force tapping mode, which enabled simultaneous acquisition of topography and multiple property maps, such as those of stiffness and adhesion. Backside Au-coated Si cantilevers (Nanosensors, FM-AuD) with a spring constant of 2–4 N/m were used; the nominal tip radius was approximately 10 nm. Prior to completing AFM measurements, we cleaned the AFM probe by using an ultraviolet light. All AFM imaging was performed at RT.

**XPS**. Measurements were performed at BL24A beamline of Taiwan Light Source (TLS) in National Synchrotron Radiation Research Center (NSRRC). The TLS is a 1.5-GeV ring operated in a top-up injection mode with a constant ring current of 360 mA. XPS measurements were probed using the incident X-ray energy of 750.0 eV and collected using the photoelectron analyzer (PHOIBOS150, SPECS). The photon energy was calibrated with the major peak of the fresh graphite sheets occurring at a BE of 284.4 eV. The averaged energy resolution was better than 0.1 eV.

**TDS**. A mass spectrometer was employed to measure the m/z ratios of the atoms or molecules desorbed from a surface when the temperature of the sample was increased. TDS experiments were conducted in a UHV chamber (Omicron VT-STM) with a base pressure of $4 \times 10^{-10}$ mbar. The sample holder, modified from a standard sample holder for Omicron VT-STM, was transferred from the UHV chamber to ambient air by using a load-lock system. Each sample was immediately mounted on the holder within a 5-min window before being moved back into the load-lock system, which was then pumped with a turbomolecular pump overnight. Subsequently, the samples were transferred into the UHV chamber. The heating process during the TDS measurements involved indirect heating by running a current through a tungsten wire, which was insulated with a ceramic tube and placed under the sample holder. The heating current was gradually increased over time during the TDS measurement. The sample temperature was measured after several TDS measurements had been conducted. The same heating procedure was used for the temperature measurement, with the temperature being measured by attaching a Chromel/Alumel thermocouple to the top side of a HOPG sample (approximately the same size as that used for the TDS measurements). The accuracy of the temperature measurements was estimated to be ±5 °C. TDS was performed using one of the three quadrupole mass spectrometers (Pfeiffer PrismaPro® QMG250, Prisma QMS 200, and Stanford Research System RGA300).

**Acknowledgements**

This research was supported by the Ministry of Science and Technology of Taiwan (MOST 106-2112-M-001-025-MY3, MOST 109-2112-M-001-048-MY3, MOST 111-2112-M-213-022, MOST 110-2112-M-032-015 and MOST 110-2112-M-032-001), the National Science and Technology Council (NSTC 112-2112-M-001-047) and Academia Sinica. The authors thank Mr. Meng-Hsuan Tsai for his practical assistance with BL20, NSRRC.

# Formation of highly stable interfacial nitrogen gas hydrate overlayers under ambient conditions

Chung-Kai Fang, Cheng-Hao Chuang, Chih-Wen Yang, Zheng-Rong Guo, Wei-Hao Hsu, Chia-Hsin Wang, and Ing-Shouh Hwang*

Institute of Physics, Academia Sinica, Nankang, Taipei 11529, Taiwan

Department of Physics, Tamkang University, Tamsui, New Taipei City 251301, Taiwan

National Synchrotron Radiation Research Center, 101 Hsin-Ann Road, Hsinchu 300092, Taiwan

Correspondence and requests for materials should be addressed to I.-S.H. (email: ishwang@phys.sinica.edu.tw).

Index of the Supplementary Information:

1. Supplementary Notes
2. Supplementary Figures

# Supplementary Notes

1. **Determination of the molecular ratio of stripe structures**

    Fig. S3 depicts the $O_{1s}$ and $N_{1s}$ XPS spectra of individual chemical binding in $HOPG_{water}$ and $HOPG_{nanoGOs}$. The measured spectra are indicated with green lines, and the overall fitting curves of these spectra are denoted by black dotted lines. The other colored lines represent the fitting curves of each component. CasaXPS software was used for the quantification and curve fitting of the XPS spectra. The curve fitting and simulation of the spectra involved Shirley background subtraction and a mixing Gaussian–Lorentzian curve (G/L ratio 70/30) peak shape. The $O_{1s}$ spectra for $HOPG_{water}$ was assumed to comprise two components: one at a BE of 532.4 eV (red line in Fig. S3a) that was associated with the chemisorption of oxygen on HOPG, which was derived from the $O_{1s}$ spectra for $HOPG_{fresh}$ (Fig. 5a), and the other associated with water molecules. Our fitting analysis revealed that a fitting curve with a BE of 533.2 eV for water molecules (blue line in Fig. S3a) yielded a fitting curve (black dotted line in Fig. S3a) that aligns well with the measured spectra (green line in Fig. S1a). The BE of 533.2 eV is consistent with those reported in previous studies for water ice on solid substrates (532.8–533.4 eV)[1–2]. To fit the measured $O_{1s}$ spectra of $HOPG_{nanoGOs}$ (Fig. S3b), we included three fitting peaks at BE of 531.6 eV (purple), 532.4 eV (dark yellow), and 533.2 eV (navy), which corresponded to C=O, C-OH, and C-OOH in graphene oxide[3], respectively, in addition to the fitting curves at BEs of 532.2 eV (chemisorption of oxygen) and 533.2 eV (water molecules). The peak associated with water molecules (533.2 eV, blue line) dominated the $O_{1s}$ spectra.

    The measured $N_{1s}$ spectra for $HOPG_{water}$ (Fig. S3c) and $HOPG_{nanoGOs}$ (Fig. S3d) can be decomposed into two peaks with BEs of 400.0±0.1eV and 401.5±0.2 eV.

Previous XPS studies examining the adsorption of $N_2$ on $TiO_2$ fiber[4] and Cr/W(110)[5] reported peaks with BEs of 400.1 and 400.0 eV, respectively. Thus the peak of 400.0±0.1eV presented in Fig. S3 c, d was attributed to $N_2$ molecules in the stripe structure. Asymmetric curves in the N 1s peak indicate the presence of another component, which can be fitted with a peak at a BE of 401.5 ± 0.2 eV. Previous studies on the N–H bond in ammonia and amino groups have revealed a peak within the range of 400 to 403 eV in $N_{1s}$ spectra[6-8]. The BE of 401.5 eV (represented by a dark yellowish-green line) might correspond to the interaction between $N_2$ molecules and the surrounding hydrogen atoms of water molecules.

To determine the molecular ratio of $H_2O$ ($X_{H_2O}$) and $N_2$ ($X_{N_2}$) in stripe structures on $HOPG_{water}$ and $HOPG_{nanoGOs}$, we used CasaXPS software to calculate the XPS peak area with a BE of 533.2 eV for $H_2O$ ($I_{H_2O}$, blue lines in Fig. S3 a, b) and the entire $N_{1S}$ area between 398 and 403 eV for $N_2$ ($I_{N_2}$, Fig. S3 c, d). The molecular ratios of $H_2O$ and $N_2$ were determined using Eq. (1) and Eq. (2), respectively[9]. The ratio of the sensitivity factor $S_O/S_N$ for oxygen relative to nitrogen was 1.62, a value supplied by the manufacturer of the analyzer (SPECS). After computing data from 13 independent XPS measurements of $HOPG_{water}$ and 14 independent XPS measurements of $HOPG_{nanoGOs}$, we determined that the stripe structures (nitrogen gas hydrate overlayer) on $HOPG_{water}$ were composed of 90±4% $H_2O$ and 10±4% $N_2$, and that those on $HOPG_{nanoGOs}$ were composed of 93±3% $H_2O$ and 7±3% $N_2$.

$$X_{H_2O} = \frac{I_{H_2O}/S_O}{I_{H_2O}/S_O + I_{N_2}/S_N} \qquad \ldots\ldots(1)$$

$$X_{N_2} = \frac{I_{N_2}/S_N}{I_{H_2O}/S_O + I_{N_2}/S_N} \qquad \ldots\ldots(2)$$

**[References]**

2. **Small percentage of D$_2$O in TDS**

We used D$_2$O with 99.9 atom% D to prepare stripe structures of HOPG$_{water}$. The smaller percentage at m/z=20 relative to m/z of 18 and 19 (Fig. 6c) was mainly due to the hydrogen-deuterium exchange during the mounting of the sample to the holder before it was transferred to the UHV chamber, when the HOPG$_{water}$ sample was exposed to ambient air (humidity of 60% to 80%) for several minutes.

**Supplementary Figures**

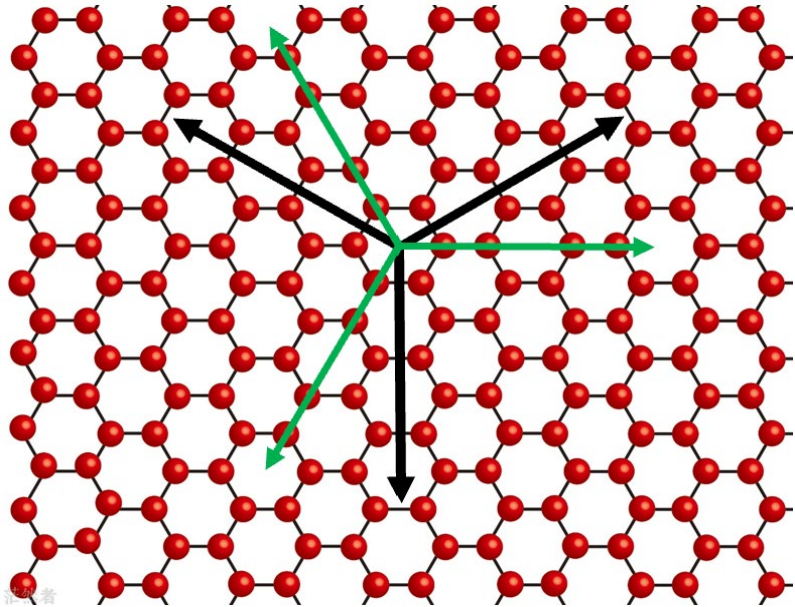

**Fig. S1 Zig-zag and arm-chair directions on the top-layer HOPG (or graphene) lattice.** Because of the three-fold symmetry of the substrate, three equivalent zig–zag directions (thick black arrows) and three arm-chair directions (green arrows) were noted. The zig-zag directions also represent the lattice directions of the substrate.

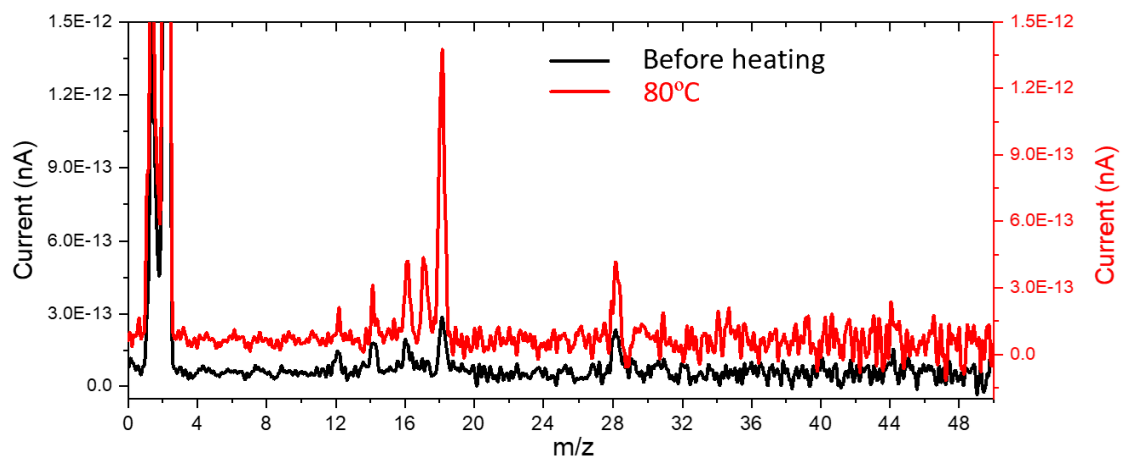

**Fig. S2 Mass spectroscopy measurements of a HOPG$_{water}$ sample before heating (black curve) and during heating at 80°C (red curve) with m/z of 1 to 50.** For clarity, the red curve is shifted upward slightly. The measurements were performed using Pfeiffer Vacuum Prisma QMS 200. The desorption of water molecules led to an increase in values at m/z of 18, 17, 16, 2, and 1. Another prominent desorption peak was observed at m/z of 28, which can be attributed to $N_2$ or CO. Desorption of $N_2$ molecules led to an increase in values at m/z of 28 and 14, which was observable through TDS. Desorption of CO should lead to increase in values at m/z of 28, 16, and 12. The increase at m/z of 12 is nonsignificant, indicating that the desorption of CO plays a less crucial role than that of $N_2$. Desorption at m/z of 15 is related to hydrocarbons; the small peak during heating indicates desorption of hydrocarbon molecules plays minor role.

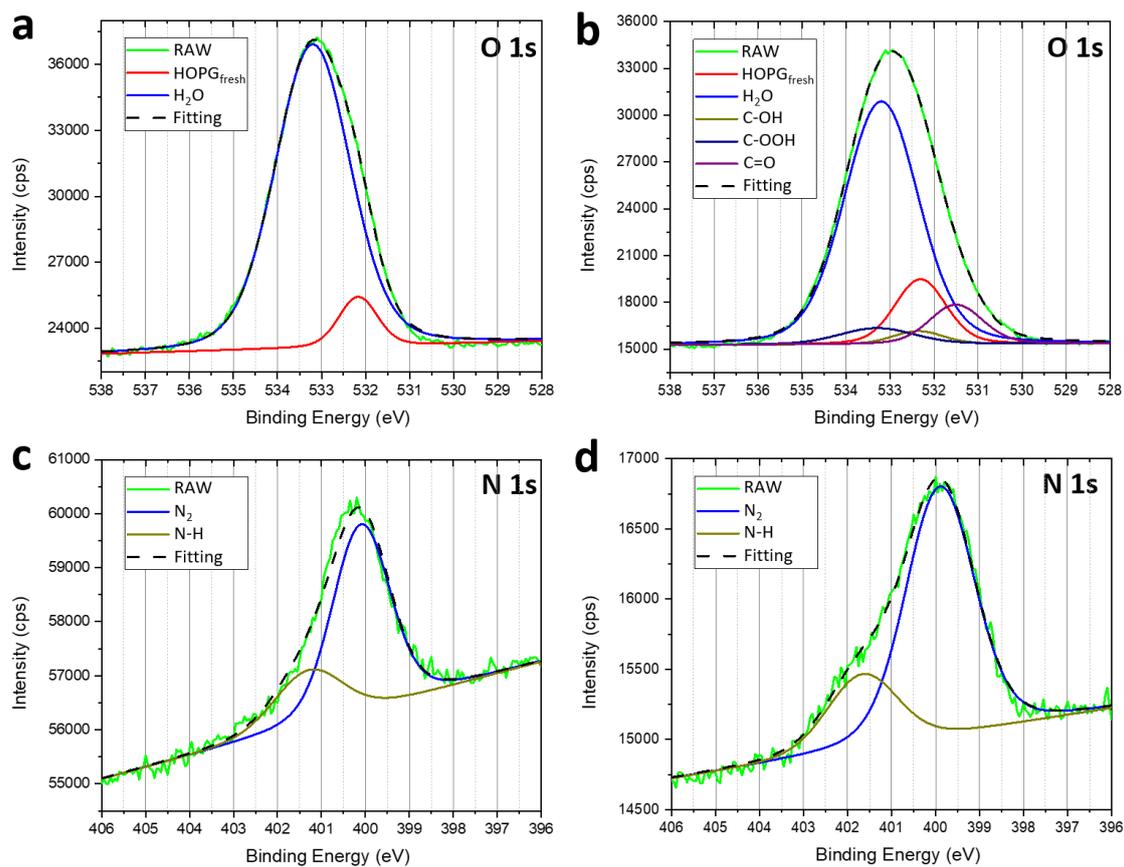

**Fig. S3 Representative $O_{1s}$ and $N_{1s}$ XPS spectra for $HOPG_{water}$ and $HOPG_{nanoGOs}$ and related fitting curves**. The measured spectra are presented using green lines, and the black dotted lines denote the overall fitting curves of the measured spectra. Other color lines represent the fitting curves of each component (inset). **a** $O_{1s}$ spectra for $HOPG_{water}$. **b** $O_{1s}$ spectra for $HOPG_{nanoGOs}$. **c** $N_{1s}$ spectra for $HOPG_{water}$. **d** $N_{1s}$ spectra for $HOPG_{nanoGOs}$.